\documentclass[10pt]{article}
\usepackage[OE]{express}

\begin{document}

\title{Exact localization length for $s$-polarized electromagnetic waves incident at the critical angle on a randomly-stratified dielectric medium}

\author{Kihong Kim\authormark{*}}

\address{Department of Energy Systems
Research and Department of Physics, Ajou University, Suwon 16499, Korea}

\email{\authormark{*}khkim@ajou.ac.kr}

\begin{abstract}
The interplay between Anderson localization and total internal reflection of electromagnetic waves incident near the critical
angle on randomly-stratified dielectric media is investigated theoretically. Using an exact analytical formula for the localization length
for the Schr\"odinger equation with a Gaussian $\delta$-correlated random potential in one dimension,
we show that when the incident angle is equal to the critical angle, the localization
length for an incident $s$ wave of wavelength $\lambda$ is directly proportional to $\lambda^{4/3}$ throughout the entire range of the wavelength,
for any value of the disorder strength.
This result is different from that of a recent study reporting that the localization
length at the critical incident angle for a binary multilayer system with random thickness variations is proportional to $\lambda$ in the
large $\lambda$ region. We also discuss the characteristic behaviors of the localization length
or the tunneling decay length for all other incident angles. Our results are confirmed by an independent numerical calculation based on the invariant imbedding method.
\end{abstract}


\section{Introduction}

Despite of the intensive studies for more than five decades,
Anderson localization of quantum particles and classical waves continues
to fascinate many researchers \cite{1,2,3,4}. Recent studies have identified new systems displaying Anderson
localization and revealed new aspects of the phenomenon \cite{5,6,7,8}.
Localization of noninteracting quantum particles in a random potential in one dimension is an old problem that has been studied
extensively for a long time \cite{9,10}. Among many theoretical results, we revisit an exact analytical formula for the localization length in the presence of
a Gaussian $\delta$-correlated random potential derived long ago by several authors \cite{11,12,13,14}.
This formula can be applied to the cases where both the strength of the random potential and the ratio between the average value of the potential and the particle energy are arbitrary.

Due to the similarity between the Schr\"odinger equation and the electromagnetic wave equation for $s$-polarized waves,
the analytical formula for the localization length mentioned above can also be applied to $s$ waves
incident obliquely on randomly-stratified (or randomly-layered) dielectric media. When the disorder-averaged value of the dielectric permittivity is smaller than the permittivity in
the incident region, a modified total internal reflection phenomenon occurs near and above the critical angle \cite{15,16}.
In this paper, we are especially interested in the interplay between Anderson localization and total internal reflection near the critical
angle, which has drawn some renewed theoretical and experimental interest recently \cite{17,18}. In particular, we show that when the incident angle is precisely equal to the critical angle, the localization
length for an incident $s$ wave of wavelength $\lambda$ is directly proportional to $\lambda^{4/3}$ throughout the entire range of the wavelength.
This result is different from that of a recent theoretical study based on the transfer matrix method reporting that the localization
length at the critical incident angle for a binary multilayer system with random layer thicknesses of deeply subwavelength scale drawn from
a uniform distribution, which has short-range
disorder correlations embedded in it, is proportional to $\lambda$ in the
large $\lambda$ region \cite{17}. In addition, we discuss the agreement and the disagreement between our results and
those of two previous works \cite{16,19}. The possible effects of disorder
correlations on localization phenomena are also discussed.
Our results are confirmed by an independent numerical calculation based on the invariant imbedding method.

\section{Model}

We are interested in the propagation and the Anderson localization of an $s$-polarized plane electromagnetic wave of
frequency $\omega$ and vacuum wave number $k_0$ ($=\omega/c$) in isotropic random dielectric media. The wave is incident obliquely on a
stratified random medium, where the dielectric permittivity $\epsilon$
varies randomly only in the $z$ direction. We assume that the
random medium lies in $0 \le z \le L$ and the wave propagates in the
$xz$ plane. In the $s$ wave case, then, the complex amplitude of the
electric field, ${\mathcal
E}$, satisfies
\begin{equation}
{{d^2{\mathcal
E}}\over{dz^2}}
+\left[{k_0}^2\epsilon(z)-q^2\right]{\mathcal
E}=0, \label{eq:s}
\end{equation}
where $q$ is the $x$ component of the wave vector.

We assume that the wave is incident from the region where
$\epsilon=\epsilon_1$ and $z>L$ and transmitted to the region
where $\epsilon=\epsilon_1$ and $z<0$.
When $\theta$ is
the angle of incidence, the quantity $q$ is equal to $k\sin\theta$, where $k=\sqrt{\epsilon_1} k_0$.
In the inhomogeneous slab located in $0\le z\le L$, $\epsilon$ is given by
\begin{eqnarray}
\epsilon=\langle\epsilon\rangle+\delta\epsilon(z),
\end{eqnarray}
where $\langle\epsilon\rangle$ is the disorder-averaged value of $\epsilon$ and $\delta\epsilon(z)$ is a
$\delta$-correlated Gaussian random function with zero average. We assume that $\langle\epsilon\rangle$ is independent of $z$.

It is convenient to introduce a normalized variable $\tilde\epsilon$ defined by
\begin{eqnarray}
{\tilde\epsilon}=\frac{\epsilon}{\epsilon_1}=1+a+\delta{\tilde\epsilon}(z),
\end{eqnarray}
where
\begin{eqnarray}
a=\frac{\langle\epsilon\rangle}{\epsilon_1}-1,
~~\delta{\tilde\epsilon}(z)=\frac{\delta\epsilon(z)}{\epsilon_1}.
\end{eqnarray}
The random variable $\delta{\tilde\epsilon}$ satisfies
\begin{eqnarray}
\langle \delta{\tilde\epsilon}(z)\delta{\tilde\epsilon}(z^\prime) \rangle &=& 2
g_0\delta(z-z^\prime),~~~\langle\delta{\tilde\epsilon}(z)\rangle=0,
\label{eq:rand1}
\end{eqnarray}
where $g_0$ is
a parameter measuring the strength of disorder and has the dimension of a length.

With some simple manipulation,
we can rewrite the wave equation in the form
\begin{eqnarray}
{{d^2{\mathcal
E}}\over{dz^2}}+\left(k\cos\theta\right)^2\left[1+\frac{a+\delta{\tilde\epsilon}(z)}{\cos^2\theta}\right]{\mathcal E}=0.
\label{eq:sche}
\end{eqnarray}
This equation has
precisely the same form as the Schr{\"o}dinger equation for
a quantum-mechanical particle in a random potential in one dimension
\begin{equation}
{{d^2\psi}\over{dz^2}}+\kappa^2\left[1-{{V(z)}\over E}\right]\psi=0,
\label{eq:sch}
\end{equation}
where $E$ ($=\hbar^2 \kappa^2/2m_0$) is the energy of the incoming particle
of mass $m_0$ and $V(z)$ is the potential. We assume that the
potential is equal to 0 for $z<0$ and $z>L$ and $V(z)=V_0+\delta
V(z)$ for $0 \le z \le L$, where $V_0$ is a real constant and
$\delta V(z)$ is a Gaussian random function with zero mean and a
white-noise spectrum:
\begin{equation}
\langle \delta V(z)\delta V(z^\prime) \rangle =
2D\delta(z-z^\prime),~~~\langle\delta V(z)\rangle=0.
\end{equation}
The correspondences between Eq.~(\ref{eq:sche}) and Eq.~(\ref{eq:sch}) are summarized in Table~\ref{tab1}.

\begin{table}[htbp]
\centering
\caption{\label{tab1} Correspondence between the electromagnetic wave equation for
obliquely incident $s$ waves [Eq.~(6)] and the Schr{\"o}dinger equation in one dimension [Eq.~(7)].}
\begin{tabular}{|c|c|}
\hline
$s$ wave equation & Schr{\"o}dinger equation \\ \hline
${\mathcal E}$ & $\psi$ \\ \hline
$k\cos\theta$ & $\kappa$ ($=\sqrt{2m_0 E}/\hbar$) \\ \hline
$a/\cos^2\theta$ & $-V_0/E$ \\ \hline
$\delta\tilde\epsilon(z)/\cos^2\theta$ & $-\delta V(z)/E$ \\ \hline
$g_0/\cos^4\theta$ & $D/E^2$ \\ \hline
\end{tabular}
\end{table}

The tunneling of a quantum-mechanical particle occurs when the particle energy $E$ ($>0$) is
smaller than $V_0$. This condition corresponds to $-a>\cos^2\theta$, or equivalently, $\sin\theta > \sqrt{{\langle\epsilon\rangle}/\epsilon_1}$,
which gives the usual expression for the critical angle, $\sin\theta_c=\sqrt{{\langle\epsilon\rangle}/\epsilon_1}$, when $0<{\langle\epsilon\rangle}<\epsilon_1$.

\section{Lyapunov exponent}
\label{sec3}

We consider the transmission of a quantum particle through the random potential $V(z)$.
We assume that the particle, which is described by a plane wave of
unit magnitude $\psi(z)=\exp\left[i\kappa(L-z)\right]$, is incident onto the barrier
from the region where $z>L$ and transmitted to the region where
$z<0$. The complex reflection
and transmission coefficients, $r=r(L)$ and $t=t(L)$, are defined by the
wave functions outside the random region:
\begin{eqnarray}
\psi(z)&=&\left\{ \begin{array}{ll}
e^{i\kappa(L-z)}+r(L)e^{i\kappa(z-L)},  &  ~z>L \\
t(L)e^{-i\kappa z},  &  ~z<0  \end{array} \right..
\label{eq:rtc}
\end{eqnarray}

We are interested in the Lyapunov exponent $\gamma$, which can be defined by
\begin{equation}
\lim_{L\rightarrow\infty}\frac{\langle \ln T\rangle}{2L}=-\gamma,
\label{eq:le}
\end{equation}
in terms of the transmittance $T$ ($=\vert t\vert^2$).
When $E$ is greater than $V_0$, $\gamma$ can be related to the the localization length $\xi$ by
\begin{equation}
\gamma=\frac{1}{2\xi}.
\end{equation}
In the tunneling situation where $E<V_0$, $\xi$ is more appropriately called as the tunneling decay length.

The exact analytical expression for $\gamma$ has been derived previously by several authors and takes the form \cite{14}
\begin{eqnarray}
\gamma=\tilde D^{1/3}F(X),
\end{eqnarray}
where
\begin{eqnarray}
\tilde D=\frac{\kappa^4 D}{E^2},~~X=\left(\frac{E^2}{\kappa D}\right)^{2/3}\left(\frac{V_0}{E}-1\right),
\end{eqnarray}
and the function $F(X)$ is expressed in terms of the Airy functions Ai and Bi and their derivatives (denoted by primes) as
\begin{eqnarray}
F(X)=\frac{{\rm Ai}(X){\rm Ai}^\prime(X)+{\rm Bi}(X){\rm Bi}^\prime(X)}{\left[{\rm Ai}(X)\right]^2+\left[{\rm Bi}(X)\right]^2}.
\end{eqnarray}
Using this and the correspondences shown in Table~\ref{tab1}, we derive the analytical expression of $\xi$ for $s$ waves
propagating obliquely in randomly-stratified dielectric media, which takes the form
\begin{eqnarray}
\frac{1}{\xi}=2k\left(k g_0\right)^{1/3}F\left(\frac{\sin^2\theta-\frac{{\langle\epsilon\rangle}}{\epsilon_1}}{\left(k g_0\right)^{2/3}}\right).
\label{eq:halp}
\end{eqnarray}

When $\langle\epsilon\rangle$ is smaller than $\epsilon_1$, there exists a critical angle $\theta_c$, at which the argument of $F$ in Eq.~(15) vanishes. Then we find that for $s$ waves incident at the critical angle, $\xi_c$ [$=\xi(\theta=\theta_c)$] is precisely given by
\begin{eqnarray}
\frac{1}{\xi_c}=2F(0)k^{4/3}g_0^{1/3}\propto \omega^{4/3}
\label{eq:linc0}
\end{eqnarray}
in the {\it entire} range of frequencies.
We can rewrite this expression in terms of the wavelength in the incident region, $\lambda$ ($=2\pi/k$), as
\begin{eqnarray}
\xi_c&=&\frac{1}{2(2\pi)^{4/3}F(0)}g_0^{-1/3}{\lambda}^{4/3}\nonumber\\
&\approx & 0.1183~ g_0^{-1/3}{\lambda}^{4/3} \propto {\lambda}^{4/3},
\label{eq:linc}
\end{eqnarray}
where we have inserted the numerical value of $F(0)$ ($\approx 0.364506$).
In Sec.~\ref{sec_iim}, we will introduce a {\it dimensionless} disorder parameter $g$ ($=kg_0/2$), in terms of which we have
\begin{eqnarray}
\frac{1}{k\xi_c}=2^{4/3}F(0)g^{1/3}\approx 0.9185~g^{1/3},
\label{eq:linb}
\end{eqnarray}
for all values of $g$ ($>0$). We emphasize that {\it not only the exponent but also the proportionality constant is universal} in this equation.

The pure power law of Eq.~(\ref{eq:linc0}) is due to the use of the $\delta$-correlated model. If the disorder has a finite correlation length,
then the validity of this power law will be restricted to a low-frequency or weak-disorder regime. In the context of the Schr\"odinger
equation, Eq.~(\ref{eq:sch}), the analogue of this law has been shown to hold over the entire energy range for a $\delta$-correlated random potential \cite{12}. In the tight-binding model, where the disorder has a finite correlation length, it has been shown that the power law
only describes the weak-disorder regime \cite{13}.

When $\theta$ is smaller than $\theta_c$ and $g$ is sufficiently small, the argument of $F$ in Eq.~(15)
approaches to large negative values.
Then, from the property of $F$ such that
\begin{eqnarray}
\lim_{X\rightarrow -\infty}F(X)=-\frac{1}{4X},
\end{eqnarray}
we obtain
\begin{eqnarray}
\ln\left(\frac{1}{k\xi}\right)\approx \ln g -\ln\left(a+\cos^2\theta\right)
\label{eq:linb3}
\end{eqnarray}
and
\begin{eqnarray}
\xi\propto k^{-2} \propto \omega^{-2} \propto \lambda^{2},
\label{eq:linc3}
\end{eqnarray}
in the weak disorder or small frequency limit. If we include the prefactor, this relation can be written as
\begin{equation}
\xi \approx \frac{a+\cos^2\theta}{2\pi^2 g_0} \lambda^{2}.
\label{eq:linc2}
\end{equation}

In Ref.~16, the localization of $s$ waves in a short-range-correlated multilayer system with a Gaussian random
distribution of the layer refractive indices has been studied using the transfer matrix method. It has been shown that the localization length at the critical angle, $\xi_c$, is related to
the width of the distribution of the refractive index, $\zeta$, by
\begin{eqnarray}
\xi_c\propto \zeta^{-2/3},
\label{eq:bd}
\end{eqnarray}
in the long-wavelength region.
It has also been shown that, when the wave is incident normally, $\xi$ satisfies
\begin{eqnarray}
\xi\propto \zeta^{-2}
\label{eq:bd2}
\end{eqnarray}
in the long-wavelength region. This result can be extended to all incident angles below $\theta_c$.
The randomly fluctuating part of the refractive index, $\delta n$, is proportional to that of the dielectric permittivity, $\delta \tilde\epsilon$, and therefore
we have $\langle \delta n(z)\delta n(z^\prime)\rangle\propto \langle \delta \tilde\epsilon(z)\delta \tilde\epsilon(z^\prime)\rangle$.
Since the coefficient of the $\delta$ function, $2g_0$, in Eq.~(5) is proportional to the width of the corresponding Gaussian probability
distribution, we conclude that
our disorder parameter $g_0$ is proportional to $\zeta^2$.
Then the scaling relations, Eqs.~(\ref{eq:bd}) and (\ref{eq:bd2}), are consistent with Eqs.~(\ref{eq:linc}) and (\ref{eq:linc2}), from which it follows that $\xi_c\propto g_0^{-1/3}$ and $\xi\propto g_0^{-1}$ respectively. Since $g_0$ should appear as the dimensionless combination $kg_0$ and $\xi$ has the dimension of a length, we also obtain $\xi_c \propto \lambda^{4/3}$ when $\theta=\theta_c$ and $\xi \propto \lambda^{2}$ when $\theta<\theta_c$.
In the long-wavelength region, these results agree precisely with ours obtained using the $\delta$-correlated model.

\section{Influence of the correlation of disorder}

The $\delta$-correlated model studied here has no correlations, while the multilayer models used in Refs.~16 and 17 have
short-range correlations embedded in them. The influence of correlations on localization has been a major topic of research
in recent years \cite{6,20}. It has been shown that long-range correlations generally have dramatic effects on localization [21--23], while
short-range correlations, except for in some special cases \cite{24,25}, have weaker effects on it.

The multilayer model of Ref.~17, which is a kind of
periodic-on-average structure, has been designed specifically to study the influence of short-range correlations on localization experimentally \cite{18},
and its application to other wave systems is limited. On the other hand, the $\delta$-correlated model is more generic
and is valuable in that it can be applied to all wave systems to study their long-wavelength properties.

Properties of Gaussian random systems can be characterized by the spectral density, $S(k)$,
which is the spatial Fourier transform of the disorder correlation function such as $\langle \delta\tilde\epsilon(z) \delta\tilde\epsilon(z^\prime)\rangle$.
For the $\delta$-correlated model given by Eq.~(\ref{eq:rand1}), we obtain
\begin{equation}
S(k)=2g_0,
\end{equation}
which is a constant independent of $k$.
The feature that $S(k)$ is finite in the $k\rightarrow\infty$ limit and $\int S(k)dk$ diverges is unphysical,
and therefore the prediction of this model in the short-wavelength region should not be trusted.

The simplest kind of a short-range-correlated model, which can be treated analytically, is given by
\begin{equation}
\langle \delta\tilde\epsilon(z) \delta\tilde\epsilon(z^\prime)\rangle=\sigma^2 \exp\left(-\vert z-z^\prime\vert/l_c\right),
\label{eq:src}
\end{equation}
where $l_c$ is the disorder correlation length. In the limit where $\sigma\rightarrow\infty$, $l_c\rightarrow 0$ and $\sigma^2 l_c\rightarrow g_0$,
this correlation function reduces precisely to Eq.~(\ref{eq:rand1}).
Its spectral density is given by
\begin{equation}
S(k)=\frac{2\sigma^2 l_c}{1+(kl_c)^2}.
\label{eq:sd2}
\end{equation}
In the long-wavelength limit ($k\rightarrow 0$), this function approaches to the constant $2\sigma^2 l_c$.

The power-law dependences of the localization length on $\lambda$ in Eqs.~(\ref{eq:linc}) and (\ref{eq:linc2})
in the large $\lambda$ limit
can be considered as examples of critical phenomena and the exponents 4/3 and 2 are the associated
critical exponents \cite{4}.
The renormalization group (RG) theory is the fundamental theory
of critical phenomena, which provides the physical basis of the universality of the macroscopic critical properties such as the critical exponent \cite{26}.
The RG theory of critical phenomena in random systems with short-range and long-range correlated disorder is a well-established subject developed a long time ago [27--29].

Under the RG transformation \cite{26}, it can be shown that the $k$-dependent term in Eq.~(\ref{eq:sd2}) is irrelevant in the macroscopic limit and does not
contribute to determining the critical exponent.
Therefore, in the long-wavelength regime, the critical exponents for the two models given by Eqs.~(\ref{eq:rand1}) and (\ref{eq:src}) will be the same, while, in the short-wavelength regime, these models will behave differently.

Somewhat more general short-range-correlated models are expected to have the spectral density of the series form
\begin{equation}
S(k)=\sum_{n=0}^{\infty}a_n k^{2n}.
\end{equation}
Under the RG transformation, all terms with $n> 0$ can be shown to flow to zero and only the leading constant term $a_0$ is relevant.
Therefore the critical exponents for this class of models will also be the same as those for the $\delta$-correlated model.
This may be the reason why the exponents for the model in Ref.~16 are the same as our exponents for both $\theta=\theta_c$ and $\theta<\theta_c$.
Using the RG theory terminology, these two parameter regions are described by two different RG fixed points.
We note that $\delta$-correlated models have been routinely used in a large number of studies on static critical phenomena in random systems \cite{28,29}
and also in those on dynamic critical phenomena driven by $\delta$-correlated noise \cite{30}.

There can be exceptions to the argument given above. They are usually related to the presence of a hidden symmetry such as the gauge symmetry.
In the literature, there are examples of low-dimensional wave systems, which include dimer type models \cite{24} and multilayer models containing alternate positive index and negative index layers \cite{25}, where the long-wavelength scaling behavior is nontrivially affected by
{\it short-range} correlations. These models may have a symmetry that constrains to change the long-wavelength
behavior of the spectral density. More specifically, a possible scenario to allow a system with short-range correlations to behave differently from the $\delta$-correlated model is that it
has a special symmetry that forces the constant term $a_0$ to remain
strictly zero under the RG transformation and $S(k)$ contains a nonanalytic leading term of the form $\vert k\vert^\beta$ with $0<\beta<2$. An argument
of a similar kind has been used previously in an RG study of a system with a gauge symmetry \cite{fg}. In Ref.~17, the authors have reported the results of analytical and numerical analysis showing that $\xi_c\propto \lambda$ at $\theta=\theta_c$. Does this result suggest that the model in Ref.~17 has a relevant perturbation or a hidden symmetry that makes its universality class
different from that of Ref.~16?
We consider this to be a very interesting question which deserves further investigation.

Long-range correlations have much more dramatic effects on critical phenomena. For instance, the model in Ref.~21 has the spectral density scaling as $S(k)\propto k^{-p}$ in the small $k$ limit ($p>0$).
Under the RG transformation, this term is more relevant than the constant term and dominates the macroscopic critical behavior.
The critical exponents will depend on the value of $p$ continuously and a nontrivial phase transition can occur in such systems.
It is a very interesting problem to study the influence of long-range correlations on the scaling behavior considered in this paper.

\section{Invariant imbedding method}
\label{sec_iim}

The reflection and transmission coefficients for $s$ waves are defined similarly to Eq.~(\ref{eq:rtc}).
For an $s$ wave of unit magnitude $\tilde
{\mathcal E}(x,z)={\mathcal E}(z)\exp\left(iqx\right)=\exp\left[ip(L-z)+iqx\right]$, where $p=k\cos\theta$, incident
on the medium, they are defined by
\begin{eqnarray}
\tilde {\mathcal E}(x,z)=\left\{ \begin{array}{ll}
\left[e^{ip(L-z)}+re^{ip(z-L)}\right]e^{iqx},    &z>L \\
te^{-ipz+iqx},    &z<0  \end{array} \right..
\end{eqnarray}
Using the invariant imbedding method \cite{31,32,33,34,35},
we derive exact differential
equations satisfied by $r$ and $t$:
\begin{eqnarray}
{{dr}\over{dl}}=2ik\cos\theta~ r(l) +{{ik}\over
{2\cos\theta}}\left[a+\delta\tilde\epsilon(l)
\right]\left[1+r(l)\right]^2,\nonumber\\
{{dt}\over{dl}}=ik\cos\theta~ t(l) +{{ik}\over
{2\cos\theta}}\left[a+\delta\tilde\epsilon(l)
\right]\left[1+r(l)\right]t(l), \label{eq:rrt}
\end{eqnarray}
which are integrated from $l=0$ to $l=L$ using the
initial conditions $r(0)=0$ and $t(0)=1$.

We can use Eq.~(\ref{eq:rrt}) in calculating the disorder averages of
various physical quantities consisting of $r$ and $t$.
We are mainly interested
in calculating the localization
length $\xi$ of the wave defined by
\begin{equation}
\lim_{L\rightarrow\infty}\langle \ln T\rangle=-\frac{L}{\xi}. \label{eq:ll}
\end{equation}
In order to obtain $\xi$, we need to compute $\langle \ln
T(l)\rangle$ in the $l\rightarrow \infty$ limit. The {\it nonrandom}
differential equation satisfied by $\langle \ln T\rangle$ is
obtained using Eq.~(\ref{eq:rrt}) and Novikov's formula \cite{36} and takes the form
\begin{eqnarray}
{1\over k}{{d}\over{dl}}\langle \ln
T\rangle=
-{g\over{\cos^2\theta}} -{\rm
Re}\left[\left(-i{{a}\over{\cos\theta}}
+{2g\over{\cos^2\theta}}\right)Z_1(l)
+{g\over{\cos^2\theta}}Z_2(l)\right],
\end{eqnarray}
where we have used the definitions $g=kg_0/2$ and $Z_n=\langle r^n\rangle$ ($n=1,2$). From this, we obtain
\begin{eqnarray}
{1\over {k\xi}}=
{g\over{\cos^2\theta}} +{\rm
Re}\left[\left(-i{{a}\over{\cos\theta}}
+{2g\over{\cos^2\theta}}\right)Z_1(l\rightarrow\infty)
+{g\over{\cos^2\theta}}Z_2(l\rightarrow\infty)\right].
\end{eqnarray}
Therefore the problem of calculating $\xi$ reduces to that of calculating $Z_1$ and $Z_2$
in the $l\rightarrow\infty$ limit.

It turns out that in order to obtain $Z_1$ and $Z_2$, we need to compute $Z_n$ for all
positive integers $n$.
An infinite number of coupled nonrandom differential
equations satisfied by $Z_n$'s are obtained using Eq.~(\ref{eq:rrt}) and Novikov's
formula and have the form
\begin{eqnarray}
&&{1\over k}{{d}\over{dl}}Z_n= \left[i\left(2\cos\theta+\frac{a}{\cos\theta}\right)
n -\frac{3g}{\cos^2\theta} n^2\right]
Z_{n}\nonumber\\
&&~~~+ n\left[i\frac{a}{2\cos\theta}-\frac{g}{\cos^2\theta} \left(
2n+1\right)
\right]Z_{n+1}+ n\left[i\frac{a}{2\cos\theta}-\frac{g}{\cos^2\theta}
\left(2n-1\right)
\right]Z_{n-1}\nonumber\\
&&~~~- \frac{g}{2\cos^2\theta}n(n+1) Z_{n+2} -\frac{g}{2\cos^2\theta}n(n-1)Z_{n-2}, \label{eq:z}
\end{eqnarray}
where $Z_{n}$ satisfies the initial
condition
$Z_{n}(l=0)=0$.

The magnitude of $Z_{n}$
decays (more rapidly for the larger value of $g$) as $n$ increases. Based on
this observation, we solve the infinite number of coupled
equations using a simple
truncation method \cite{37}. In the $l\rightarrow\infty$ limit, the left-hand side of Eq.~(\ref{eq:z}) vanishes and
we get a set of {\it algebraic} equations. This set of algebraic equations takes the form of a five-term linear recursion relation
and can provide another way to derive Eq.~(\ref{eq:halp}).

\section{Numerical results}
\label{sec4}

\begin{figure}[htbp]
\centering\includegraphics[width=10cm]{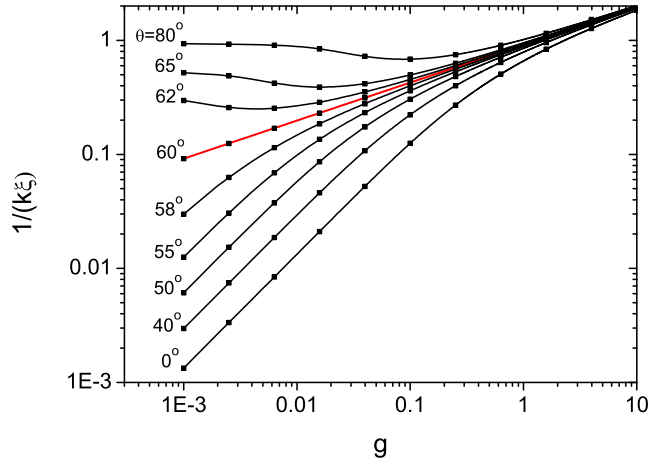}
\caption{Normalized inverse localization length (or tunneling decay length if $\theta>\theta_c$), $1/(k\xi)$, versus dimensionless disorder parameter $g$ in a log-log plot
for various values of the incident angle $\theta$, when $\langle\epsilon\rangle/\epsilon_1=0.75$
and $\theta_c=60^\circ$. The curves obtained from the analytical formula,
Eq.~(\ref{eq:halp}), are compared with the numerical results obtained using the invariant imbedding method, which are designated by square dots.
\label{fig:1}}
\end{figure}

In Fig.~\ref{fig:1}, we show the inverse of the normalized localization length (or tunneling decay length if $\theta>\theta_c$)
as a function of the dimensionless
disorder parameter $g$ in a log-log plot for various values of the incident angle $\theta$, when $\langle\epsilon\rangle/\epsilon_1=0.75$
and $\theta_c=60^\circ$. The curves obtained from the analytical formula,
Eq.~(\ref{eq:halp}), are compared with the numerical results obtained using the invariant imbedding method, which are designated by square dots.
The agreements are perfect in all cases.

When $\theta$ is equal to $\theta_c$, we note that the curve (designated by the red color) is a straight line in the entire range of $g$. This is obvious from Eq.~(\ref{eq:linb}), from which we have
\begin{eqnarray}
\ln\left(\frac{1}{k\xi_c}\right)= \frac{1}{3}\ln g +{\rm constant}.
\label{eq:linb2}
\end{eqnarray}
As $g$ ($=kg_0/2$) increases to large numbers, or equivalently, as the disorder strength or the frequency increases to large values, the argument $X$ of the function $F$ in Eq.~(\ref{eq:halp}) approaches to zero. Then it follows that all curves converge to the straight line obtained when $\theta=\theta_c$
as $g$ becomes large, as demonstrated clearly in Fig.~\ref{fig:1}.

In Ref.~19, it has been claimed that the localization length for electromagnetic waves in one dimension in the high frequency
limit is proportional to $\omega^{-2/3}$, while it is proportional to $\omega^{-2}$ in the low frequency limit.
Though the low frequency result is correct, the high frequency result differs clearly from our result that $\xi\propto \omega^{-4/3}$.

When $\theta$ is smaller than $\theta_c$ and $g$ is sufficiently small, all curves approach straight lines with the slope equal to one
in the logarithmic variables. This well-known behavior is obtained from Eq.~(20), which leads to the $\xi\propto \lambda^2$ scaling in Eq.~(22).
We note that similar results were obtained for the short-range-correlated multilayer model studied in Ref.~16.

When $\theta$ is above the critical angle, the curves show a nonmonotonic behavior, namely the tunneling decay length first increases,
reaches a maximum, and then decreases, as $g$ increases from zero. This can be understood easily from
the asymptotic expansion of $F$ \cite{14}:
\begin{eqnarray}
\lim_{X\rightarrow \infty}F(X)=\sqrt{X}-\frac{1}{4X}+\cdots,
\end{eqnarray}
which leads to
\begin{eqnarray}
\frac{1}{k\xi}=2\left(\sin^2\theta-\frac{\langle\epsilon\rangle}{\epsilon_1}\right)^{1/2}
-\frac{g}{\sin^2\theta-\frac{\langle\epsilon\rangle}{\epsilon_1}}+\cdots.
\end{eqnarray}
The leading effect of a weak disorder is to enhance the tunneling decay length.
The related phenomenon of disorder-enhanced tunneling transmission has been studied previously by several authors \cite{14,38,39,40}.

\begin{figure}[htbp]
\centering\includegraphics[width=9cm]{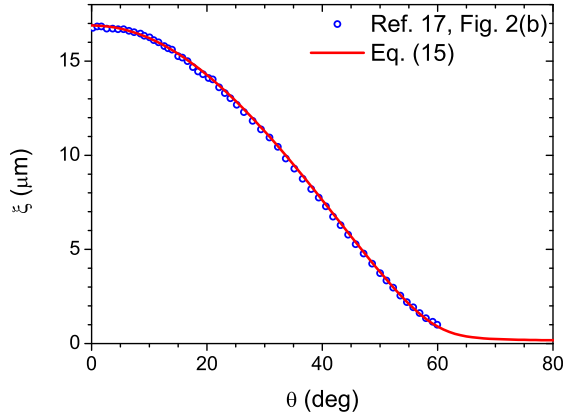}
\caption{Comparison between Fig.~2(b) of Ref.~17 and the localization length versus incident angle curve obtained from Eq.~(\ref{eq:halp}), when $\lambda=1$ $\mu$m and $g_0=0.00225$ $\mu$m.
\label{fig:2}}
\end{figure}

\begin{figure}[htbp]
\centering\includegraphics[width=9cm]{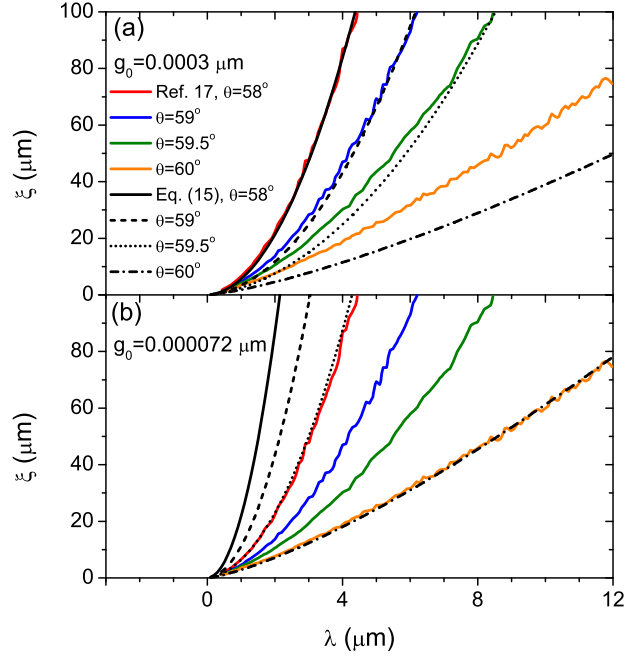}
\caption{Comparison between Fig.~2(a) of Ref.~17 (colored lines) and the localization length versus wavelength curves (black lines) obtained from Eq.~(\ref{eq:halp}), when $\theta=58^\circ$, $59^\circ$, $59.5^\circ$ and $60^\circ$, and when (a) $g_0=0.0003$ $\mu$m and (b) $g_0=0.000072$ $\mu$m.
\label{fig:3}}
\end{figure}

In Figs.~\ref{fig:2} and \ref{fig:3}, we attempt to compare the results reported in Ref.~17 with those of our calculations. We plot the localization length versus incident angle when
$\langle\epsilon\rangle/\epsilon_1=0.75$ and compare the result with Fig.~2(b) of Ref.~17, which was obtained for $\lambda=1$ $\mu$m. For the numerical comparison, we have chosen the disorder parameter $g_0=0.00225$ $\mu$m. With this choice of $g_0$, we find that the agreement is excellent.
We also find that the curve is smooth and shows no sharp feature across $\theta=\theta_c$.
This smooth behavior is a consequence of the fact that the scaling function $F(X)$ in Eq.~(15) is regular for all finite values of $X$.
This implies that the distinction between the localization length at $\theta<\theta_c$ and the tunneling decay length at $\theta>\theta_c$
is blurred in the presence of disorder.

In Fig.~\ref{fig:3}, we plot $\xi$ versus wavelength for several values of $\theta$ near $\theta_c$ and for a fixed value of $g_0$ and
compare the result with Fig.~2(a) of Ref.~17. We could not find a value of $g_0$ that gives good agreements for all considered values of $\theta$.
In Fig.~\ref{fig:3}(a), we compare our results obtained using $g_0=0.0003$ $\mu$m with Fig.~2(a) of Ref.~17.
When $\theta$ is equal to $58^\circ$ and $59^\circ$, the agreement is pretty good and when $\theta$ is $59.5^\circ$, it is not too bad. For $\theta=60^\circ$, however,
our result differs greatly from that of Ref.~17. In Fig.~\ref{fig:3}(b), we compare our results obtained using $g_0=0.000072$ $\mu$m with Fig.~2(a) of Ref.~17. For this choice of $g_0$, our result at $\theta=60^\circ$ agrees very well with that of Ref.~17, but the results for other values of $\theta$ do not agree with those of Ref.~17 at all. The fact that we have obtained very good agreements in some curves is interesting and may suggest that the two models have close similarities in these wavelength scales despite of the obvious differences on a short-range scale. The reason why
 we do not obtain good agreements in all curves using the same disorder parameter is not clear at this stage and we believe it is worthwhile to elucidate it. In Ref.~17, it has been proposed that the data points for $\theta=\theta_c=60^\circ$ should follow a straight line. However, it is interesting to notice that, in this wavelength scale, our result shown in Fig.~\ref{fig:3}(b) demonstrates that the data can be represented very well by the $\lambda^{4/3}$ dependence. It will be very helpful to compare the results of this paper with the calculations of Ref.~17 using equivalent parameters on an equal ground in a future work.

\section{Conclusion}
\label{sec5}

In this paper, we have studied the interplay between Anderson localization and total internal reflection of electromagnetic waves
incident near the critical
angle on randomly-stratified dielectric media. In particular, we have shown that when the incident angle is equal to the critical angle, the localization
length for an incident $s$ wave of wavelength $\lambda$ is directly proportional to $\lambda^{4/3}$ throughout the entire range of the wavelength.
A further theoretical investigation of the scaling behavior for various kinds of disorder correlations and experimental studies of the results reported here on randomly-layered dielectric structures are highly desired.

\section*{Funding}
National Research Foundation of Korea Grant funded by the Korean Government (NRF-2015R1A2A2A01003494).

\end{document}